\documentclass{ws-ijmpcs}

\usepackage{slashed,amsmath}
\newcommand{\Li}{\operatorname{Li}}

\def\trimmarks{}

\begin{document}

\markboth{V. M.  Braun \& A. N. Manashov}
{QCD evolution equations from conformal symmetry}

\catchline{}{}{}{}{}
%
%%%%%%%%%%%%%%%%%%%%%%%%%%%%%%%%%%%%%%%%%%%%%%%%%%%%%%%%%%%%%%%%%%%%

\title{QCD evolution equations from conformal symmetry}
%\footnote{Not more than 3 lines. Typeset the title in 10 pt roman, uppercase and boldface.}

\author{V.~M.~Braun}

\address{Regensburg University, D-93040, Regensburg, Germany\\
vladimir.braun@physik.uni-regensburg.de}

\author{A.~N.~Manashov\footnote{permanent address: 
\emph{St.-Petersburg University, 199034, St.-Petersburg, Russia} }}

\address{Regensburg University, D-93040, Regensburg, Germany\\
alexander.manashov@physik.uni-regensburg.de}

\maketitle

\begin{history}
\received{\today}
%\revised{Day Month Year}
%\published{Day Month Year}
\end{history}

\begin{abstract}
QCD evolution equations in $\text{MS}$-like schemes can be recovered from the same equations in a
modified theory, QCD in non-integer $d=4-2\epsilon$ dimensions, which 
enjoys exact scale and conformal invariance at the critical point. 
Restrictions imposed by the conformal symmetry of the modified theory 
allow one to obtain complete evolution kernels in integer (physical) dimensions 
at the given order of perturbation theory from the spectrum of 
anomalous dimensions added by the calculation of the special conformal anomaly at one order less. 
We use this technique to derive
two-loop evolution equations for flavor-nonsinglet quark-antiquark light-ray operators that encode
the scale dependence of generalized hadron parton distributions.
\end{abstract}

\ccode{PACS numbers: 12.38.Bx,11.10.Hi,11.15.Bt}

\section{Introduction}
Studies of hard exclusive reactions constitute  an important part of the research programs at all major
 accelerator facilities. The theoretical description of
such processes involves operator matrix
elements between states with different momenta, dubbed generalized parton distributions (GPDs), or vacuum-to-hadron
matrix elements, the distribution
amplitudes (DAs). Scale dependence of these distributions is governed by the renormalization group (RG) equations
for the corresponding (nonlocal) operators and are known, at present, to the two-loop
accuracy~\cite{Belitsky:1998vj,Belitsky:1999hf}.
 This is one order less compared to the RG equations for the corresponding
``inclusive'' distributions  involving forward matrix elements~\cite{Moch:2004pa,Vogt:2004mw} and closing this gap
is desirable. The direct calculation is very challenging.
Moreover, since the two-loop RGEs for GPDs are already very cumbersome,
finding a suitable representation for the results becomes part of the problem.

It has been known for a long time~\cite{Makeenko:1980bh} that 
one-loop evolution kernels can be restored from  the corresponding anomalous dimensions
thanks to conformal symmetry of the QCD Lagrangian. 
The generalization of this technique beyond the leading order was developed by D.~M\"uller~\cite{Mueller:1991gd},
who has shown that restrictions based on conformal symmetry
allows one to restore full evolution kernels at given order of perturbation theory 
from the spectrum of anomalous dimensions at the same order, and the calculation of the special conformal anomaly at one order less.
This technique was  used to calculate the two-loop evolution kernels in momentum space for the
GPDs~\cite{Belitsky:1998vj,Belitsky:1999hf,Mueller:1993hg,Belitsky:1997rh}.
In Refs.~[\refcite{Braun:2013tva,Braun:2014vba}] we suggested a different approach to achieve the same goal.
Instead of studying effects of the conformal symmetry \emph{breaking} in the 
physical theory~\cite{Mueller:1993hg,Belitsky:1997rh} it was proposed 
to make use of the \emph{exact} conformal symmetry of a modified theory -- QCD in $d=4-2\epsilon$ dimensions at critical coupling.
Exact conformal symmetry simplifies considerably the analysis and
also suggests the optimal representation for the results in terms
of light-ray operators.
We expect that this technique will become increasingly advantageous in higher orders.

This  approach was illustrated in~[\refcite{Braun:2013tva}] on several examples to the two- and
three-loop accuracy for scalar theories and used in~[\refcite{Braun:2014vba}] to obtain two-loop evolution equations
for the flavor flavor-nonsinglet light-ray operators. The applications to gauge theories involve
several subtleties that will be discussed below.

%%%%%%%%%%%%%%%%%%%%%%%%%%%%%%%%%%%%%%%%%%%%%%%%%%%%%%%%%%%%%%%%%%%%%%%%%%%%%%%%%%%%%%%%%%%%%%%%%%%%%%%%%%%%%%%%%%%%%%
\section{Preliminaries}
%%%%%%%%%%%%%%%%%%%%%%%%%%%%%%%%%%%%%%%%%%%%%%%%%%%%%%%%%%%%%%%%%%%%%%%%%%%%%%%%%%%%%%%%%%%%%%%%%%%%%%%%%%%%%%%%%%%%%%

Conformal symmetry transformations have the simplest form for the so-called light-ray operators that 
can be  understood as generating functions for the renormalized leading-twist local operators:
\begin{eqnarray}
 [\mathcal{O}](x;z_1,z_2) &\equiv& [\bar q(x+z_1n)\slashed{n} q(x+z_2n)]
~\equiv~ \sum_{m,k} \frac{z_1^m z_2^k}{m!k!} [(D_n^m\bar q)(x) \slashed{n} (D_n^k q)(x)].
\label{LRO}
\end{eqnarray}
Here $q(x)$ is a quark field, the Wilson line is implied between the quark fields on the light-cone,
$D_n = n_\mu D^\mu$ is a covariant derivative,  $n^\mu$ is an auxiliary light-like vector, $n^2=0$, that ensures
symmetrization and subtraction of traces of local operators.
The square brackets $[\ldots]$ stand for the renormalization  in the MS scheme.
We will tacitly assume that the quark and antiquark have different flavor so that there is no mixing with gluon operators.
In most situations the overall coordinate~$x$ is irrelevant and can be put to zero; we will often abbreviate
$\mathcal{O}(z_1,z_2) \equiv \mathcal{O}(0; z_1,z_2)$.

The RGE for light-ray operators  takes  the form~\cite{Balitsky:1987bk} (here and below $a=\alpha_s/4\pi$)
\begin{align}\label{RGO}
\Big(M{\partial_M}+\beta(a)\partial_{a} +\mathbb{H}(a)\Big)[\mathcal{O}(z_1,z_2)]=0\,,
\end{align}
where $\mathbb{H}$ is an integral operator acting on the quark light-cone coordinates, $z_i$. It can be written as
\begin{align}
 \mathbb{H}[\mathcal{O}](z_1,z_2) = \int_0^1 \!d\alpha\int_0^1 d\beta\, h(\alpha,\beta)\, [\mathcal{O}](z_{12}^\alpha,z_{21}^\beta)\,,
\label{hkernel}
\end{align}
where
%\begin{align}
$z_{12}^\alpha\equiv z_1\bar\alpha+z_2\alpha $, $\bar\alpha=1-\alpha$
%\end{align}
and
\begin{align}
h(\alpha,\beta)=a\,h^{(1)}(\alpha,\beta)+ a^2\,h^{(2)}(\alpha,\beta)+\ldots
\end{align}
is a certain weight function (kernel). 
%Evidently, the powers $(z_{1}-z_{2})^N$
%
One can show that the powers $[\mathcal{O}](z_1,z_2)  \mapsto (z_{1}-z_{2})^{N-1}$ 
are eigenfunctions of the  operator $\mathbb{H}$, and the corresponding eigenvalues
\begin{align}\label{hmoments}
\gamma_N=\int d\alpha d\beta\, h(\alpha,\beta)(1-\alpha-\beta)^{N-1}\,
\end{align}
are nothing else as the anomalous dimensions of local operators of spin $N$~\cite{Braun:2013tva}.
% (with $N-1$ derivatives).
%
   The kernel
  $h(\alpha,\beta)$ is a function of two variables so that the knowledge of the anomalous dimensions $\gamma_N$ is not sufficient, 
in general, to find it.
In a conformal theory, however, it is expected that the operator $\mathbb{H}$ commutes with the generators of 
the $SL(2)$ transformations, $[\mathbb{H},S_\alpha]=0$. At the leading order the generators take the canonical form
\begin{align}
S^{(0)}_+=z_1^2\partial_{z_1}+z_2^2\partial_{z_2}+2(z_1+z_2), &&
S^{(0)}_0=z_1\partial_{z_1}+z_2\partial_{z_2}+2, &&
S^{(0)}_-=-\partial_{z_1}-\partial_{z_2}\,.
\label{canon}
\end{align}
Up to the trivial case $h(\alpha,\beta)=\delta(\alpha)\delta(\beta)$ the kernel of an operator commuting  with
the  canonical generators~(\ref{canon})
is   a function of one variable only, %~\cite{Braun:1999te},
$
h (\alpha,\beta) = \bar h (\tau)\,,
$
where $\tau = {\alpha\beta}/{\bar\alpha\bar\beta}\, $ is the so-called  con\-for\-mal ratio.
The function of one variable $\bar h (\tau)$ is determined uniquely by its moments (\ref{hmoments}) and 
can  easily be reconstructed.
It turns out that the  one-loop kernel $h^{(1)}(\alpha,\beta)$ takes a remarkably simple form~\cite{Braun:1999te}
\begin{align}
      h^{(1)}(\alpha,\beta) = -4 C_F\left[\delta_+(\tau) + \theta(1-\tau)-\frac12\delta(\alpha)\delta(\beta)\right],
\label{QCD-LO}
\end{align}
where the regularized $\delta$-function, $\delta_+(\tau)$, is defined as
\begin{align}
\int d\alpha d\beta\, \delta_+(\tau)f(z_{12}^\alpha,z_{21}^\beta)&\equiv\int_0^1 d\alpha\int_0^{1} d\beta\, \delta(\tau)
\Big[f(z_{12}^\alpha,z_{21}^\beta)-f(z_1,z_2)\Big]\,.
\label{delta-function}
\end{align}

Beyond one loop, conformal symmetry in QCD is broken by quantum corrections but, nevertheless, still imposes 
nontrivial constraints. We will show that: first, it is possible to construct the operators $S_\alpha(a)$ that commute with the evolution kernel
$\mathbb{H}(a)$ in the four-dimensional interacting theory,
$[\mathbb{H}(a),S_\alpha(a)]=0$ and, second, that this property guarantees that the kernel $\mathbb{H}(a)$ can be
restored from its spectrum.
To this end we will go over to the theory in noninteger, $d=4-2\epsilon$, dimensions at the intermediate steps.

%%%%%%%%%%%%%%%%%%%%%%%%%%%%%%%%%%%%%%%%%%%%%%%%%%%%%%%%%%%%%%%%%%%%%%%%%%%%%%%%%%%%%%%%%%%%%%%%%%%%%%%%%%%%%%%%%%%%%%
\section{QCD in $d=4-2\epsilon$ dimensions}
%%%%%%%%%%%%%%%%%%%%%%%%%%%%%%%%%%%%%%%%%%%%%%%%%%%%%%%%%%%%%%%%%%%%%%%%%%%%%%%%%%%%%%%%%%%%%%%%%%%%%%%%%%%%%%%%%%%%%%
\noindent
The QCD Lagrangian in $d = 4 - 2\epsilon$ dimensional Euclidean space in covariant gauge has the
form%
%\footnote{Our discussion of critical properties of QCD follows closely Ref.~\cite{Ciuchini:1999cv}}
%
\begin{align}\label{QCD-L}
\mathcal{L}=\bar q(\slashed{\partial}-ig\slashed{A})q+\frac14 F_{\mu\nu}^a F^{a,\mu\nu}+
\partial_\mu \bar c^a(D^\mu c)^a+\frac1{2\xi}(\partial A^a)^2.
\end{align}
For large number of flavours, $n_f$, the beta function
\begin{align}
\beta(a)=M\partial_M a=2a\Big(-\epsilon - b_0 a+ \mathcal{O}(a^2)\Big)\,, && b_0 = \frac{11}{3} N_c - \frac23 n_f\,,
\end{align}
has a nontrivial zero for the finely-tuned (critical) value of the coupling $a_*=-\epsilon/b_0+O(\epsilon^2)$.
The theory  thus enjoys exact scale  and conformal
invariance\,\footnote{ QCD is critically  equivalent to the Non-Abelian Thirring model~\cite{Hasenfratz:1992jv}
that allows one to develop technique for calculation critical indices different from the standard perturbative expansion,
see e.g. Refs.~[\refcite{Gracey:1996he,Bennett:1997ch,Ciuchini:1999cv}].}
 at the critical point~\cite{Banks:1981nn,Hasenfratz:1992jv}.
As a consequence, the RGEs are exactly conformally invariant,
but the generators are  modified by quantum corrections as compared to their canonical
expressions (\ref{canon}):
\begin{align}
 S_\alpha=S_\alpha^{(0)}+a_*\,\Delta S^{(1)}_\alpha + a_*^2 \, \Delta S^{(2)}_\alpha +\ldots
\label{expandS}
\end{align}
One can show that the generator $S_-$ (translation) does not receive any corrections, $S_-=S_-^{(0)}$,
the deformation of $S_0$ can be calculated exactly in terms of the evolution operator 
(to all orders in perturbation theory)~\cite{Braun:2013tva},
whereas the deformation of $S_+$ is nontrivial and has to be calculated explicitly to the required accuracy~\cite{Braun:2014vba}:
\begin{align}
   S_0\, =& S_0^{(0)} -\epsilon+\frac12 \mathbb{H}(a_*)\,, \qquad \mathbb{H}(a_*) = a_*\,\mathbb{H}^{(1)}+a_*^2\,\mathbb{H}^{(2)}+\ldots
\label{exactS0}\\
   S_+ =& S_+^{(0)} + (z_1+z_2)\Big(-\epsilon+ \frac12 a_* \mathbb{H}^{(1)}\Big)
+  a_*(z_1-z_2)\Delta_+ + \mathcal{O}(\epsilon^2)\,,
\label{exactS}
\end{align}
where
\begin{align}
\Delta_+ [\mathcal{O}](z_1,z_2)
= -2C_F\int_0^1d\alpha\Big(\frac{\bar\alpha}\alpha+\ln\alpha\Big) \Big[[\mathcal{O}](z_{12}^\alpha,z_2)-[\mathcal{O}](z_1,z_{21}^\alpha)\Big]\,.
\label{result1}
\end{align}
From the  technical point of view this calculation replaces evaluation of the conformal
anomaly in the theory  in integer dimensions %via the conformal Ward identities (CWI)
in the approach due to D.~M{\"u}ller~\cite{Mueller:1991gd}.

The evolution kernel at the critical point has to commute with the symmetry generators,
$[S_\alpha(a_*),\mathbb{H}(a_*)]=0$. Taking into account Eq.~(\ref{exactS0})  one concludes that $\mathbb{H}(a_*)$
commutes with the two canonical generators,
$[S_-^{(0)},\mathbb{H}(a_*)]=[S_0^{(0)},\mathbb{H}(a_*)]=0$, while expanding the last
commutator in series in $a_*$
 one obtains a nested set of commutator relations~\cite{Braun:2013tva}
\begin{align}
&[S_+^{(0)},\mathbb{H}^{(1)}]=~0\,,
\notag\\
&[S_+^{(0)},\mathbb{H}^{(2)}]=~[\mathbb{H}^{(1)},\Delta S_+^{(1)}]\,,
\notag\\
&[S_+^{(0)},\mathbb{H}^{(3)}]=~[\mathbb{H}^{(1)},\Delta S_+^{(2)}]+[\mathbb{H}^{(2)},\Delta S_+^{(1)}]\,,
\label{nest}
\end{align}
etc. Note that the commutator of the canonical generator $S_+^{(0)}$ with the evolution kernel $\mathbb{H}^{(k)}$
on the l.h.s. is given in terms of the kernels $\mathbb{H}^{(m)}$ and the corrections to the generator
$\Delta S_+^{(m)}$ of order, $m < k$.
The  relations~(\ref{nest}) can be viewed as inhomogeneous first-order
differential equations on the kernels $\mathbb{H}^{(k)}$. Their solution determines $\mathbb{H}^{(k)}$ up to an
$SL(2)$-invariant term (solution of the corresponding homogeneous equation  $[\mathbb{H}_{inv}^{(k)},S_\alpha^{(0)}]=0$),
which can be restored from the spectrum of the anomalous dimensions. This procedure is described
in detail in Ref.~[\refcite{Braun:2013tva}].

Last but not least, it is well known that in $\text{MS}$-like schemes the evolution kernels (anomalous dimensions)
do not depend on the space-time dimension.  % by construction.
%Indeed, the renormalization $\mathbb{Z}$ factors relating the renormalized and bare light-ray operators
%$[\mathcal{O}](z_1,z_2) = \mathbb{Z}\,\mathcal{O}(z_1,z_2)$ are given by the expansion
%%
%\begin{align}
%\mathbb{Z}=1+\sum_{j=1}^\infty \epsilon^{-j}\sum_{k=1}^\infty  a_s^k \, \mathbb{Z}_{jk}\,,
%\end{align}
%%
%where $\mathbb{Z}_{jk}$ have the integral representation similar to (\ref{hkernel}) in terms of functions
%of two variables, $Z_{jk}(\alpha,\beta)$ that do not depend on $\epsilon$.
It means that the kernel  $\mathbb{H}(a)$ can be restored from the kernel at the critical
point, $\mathbb{H}(a_*)$, simply by replacing $a_*\to a$ in the power series for $\mathbb{H}(a_*)$.
Finally, rewriting  $\epsilon$ in terms of the critical coupling, $\epsilon = -b_0 a_s^\ast +
\mathcal{O}(a^{*2}_s)$, in the generators $S_0(a_*),\, S_+(a_*)$ one immediately concludes that the generators
$S_\alpha(a)$, commute with the kernel $\mathbb{H}(a)$.  In this way the evolution kernel in four-dimensional theory
inherits the symmetries of the evolution kernel in conformal theory.

%%%%%%%%%%%%%%%%%%%%%%%%%%%%%%%%%%%%%%%%%%%%%%%%%%%%%%%%%%%%%%%%%%%%%%%%%%%%%%%%%%%%%%%%%%%%%%%%%%%%%%%%%%%%%%%%%%%%%%
\section{Conformal Ward Identities}
%%%%%%%%%%%%%%%%%%%%%%%%%%%%%%%%%%%%%%%%%%%%%%%%%%%%%%%%%%%%%%%%%%%%%%%%%%%%%%%%%%%%%%%%%%%%%%%%%%%%%%%%%%%%%%%%%%%%%%

To begin with, action of the generators  $S_\alpha$ on the light-ray operator (which is auxiliary and scheme-dependent object)
 has to be defined in a consistent way. We will do this by  expanding the light-ray operator over local conformal operators
that can be classified according to their transformation properties with respect to the conformal group.
These are determined by the nature of the critical point and are scheme-independent 
(i.e. can be viewed as ``physical'' observables).

The transformation laws for the leading-twist operators are completely fixed by their critical dimension and spin. 
A local operator that transforms under dilatation ($\mathbf{D}$) and special conformal transformation
($\mathbf{K}^\mu$) as follows:
\begin{align}
\label{DO}
i [\mathbf{D},[\mathcal{O}_{N}](x)] &=\Big(x\partial_x+\Delta_{N}^*\Big)\,[\mathcal{O}_{N}](x)\,,
\\
\label{SKO}
i[\mathbf{K}^\mu,[\mathcal{O}_{N}](x)]&=\left[2x^\mu(x\partial)-x^2\partial^\mu+2\Delta_N^*\, x^\mu
+2x^\nu\left(n^\mu\frac{\partial}{\partial n^\nu}
-n_\nu\frac{\partial}{\partial n_\mu}\right)\right]\,[\mathcal{O}_{N}](x)\,.
\end{align}
is called a conformal operator, by definition. The light-ray operator can be expanded over the basis of conformal operators
$\mathcal{O}_{Nk}(x)=\partial_+^k[\mathcal{O}_{N}](x)$ where $\partial_+\equiv (n\partial)$ with certain coefficient functions
\begin{align} \label{expansion}
 [\mathcal{O}(x;z_1,z_2)] =
\sum_{Nk}\Psi_{Nk}(z_1,z_2) \, \mathcal{O}_{Nk}(x)\,.
\end{align}
The functions $\Psi_{Nk}$ are homogeneous polynomials of degree $N+k$ of the quark coordinates, and, 
in general, depend on the coupling $a_*$.
They can be thought of as coordinates of the light-ray operator in
the conformal basis spanned by $\mathcal{O}_{Nk}$.

Action of the conformal symmetry generators on $\mathcal{O}_{Nk}$ follows from (\ref{DO}), (\ref{SKO}) i.e. it is  
fixed by their transformation properties (scaling dimension and spin). 
For the light-ray operators, obviously,
\begin{align}
i[\mathbf{D},[\mathcal{O}(x;z_1,z_2)]]&=\sum_{Nk}\Psi_{Nk}(z_1,z_2) \, i[\mathbf{D},\mathcal{O}_{Nk}(x)]\,,
\notag\\
i[\mathbf{K}^\mu,[\mathcal{O}(x;z_1,z_2)]]&=\sum_{Nk}\Psi_{Nk}(z_1,z_2) \, i[\mathbf{K}^\mu,\mathcal{O}_{Nk}(x)]\,,
\end{align}
and similar for the other generators.
 Taking into account the expressions in~(\ref{DO}), (\ref{SKO}) one obtains after some algebra  
(recall that the operator $[\mathcal{O}(x;z_1,z_2)]$ depends implicitly on the auxiliary vector~$n$)
\begin{align}
i[(n\mathbf{P})[\mathcal{O}(x;z_1,z_2)]]&=-S_-[\mathcal{O}(x;z_1,z_2)]\,, \\
i[\mathbf{D},[\mathcal{O}(x;z_1,z_2)]]&=\Big((x\partial_x)+2S_0-n\partial_n\Big)[\mathcal{O}(x;z_1,z_2)]\,,
\\
\label{lrK}
\frac{i}{2}[\mathbf{K}^\mu,[\mathcal{O}(x;z_1,z_2)]]&=\Big(n^\mu S_+ + 2x^\mu S_0+x^\mu (x\partial_x)-\frac12 x^2\partial^\mu
\notag\\
&\quad
+n^\mu(x\partial_n)-x^\mu(n\partial_n)-(xn)\partial_n^\mu\Big)[\mathcal{O}(x;z_1,z_2)]\,.
\end{align}
where the operators $S_+$ and $S_0$ are defined by their action on the coefficient functions of conformal operators as follows
\begin{align}\label{SS+}
&S_0\Psi_{Nk}(z_1,z_2)=(j_N+k)\Psi_{Nk}(z_1,z_2)\,, 
\notag\\
&S_+\Psi_{Nk}(z_1,z_2)=(k+1)(2j_N+k)\Psi_{Nk+1}(z_1,z_2)\,.
\end{align}
Here $j_N=(\Delta_N^*+N)/2$ is the conformal spin of the operator. 
For the special choice $x=0$ in Eq.~(\ref{lrK}) one obtains 
\begin{align}
&i[\mathbf{K}^\mu,[\mathcal{O}(z_1,z_2)]]=2n^\mu S_+\,[\mathcal{O}(z_1,z_2)], 
\notag\\
&i[\mathbf{D},[\mathcal{O}(z_1,z_2)]]=(2S_0-(n\partial_n))[\mathcal{O}(z_1,z_2)]\,.
\end{align}
This definition guarantees that the generators $S_\alpha$ satisfy the $SL(2)$ commutation relations.

The expression~(\ref{exactS0}) for the generator $S_0$ follows directly from the definition~(\ref{SS+}), taking 
into account that the polynomials $\Psi_{Nk}$ are eigenfunctions of the evolution kernel,
 $\mathbb{H}(a_*)\Psi_{Nk}=\gamma_N(a_*)\Psi_{Nk}$. 
%In order to derive explicit expression for  the generator  $S_+$ we note that due to
Next, it follows from Eq.~(\ref{lrK}) that the correlation function  of two nonlocal operators 
defined with respect to different auxiliary vectors, $n$ and $\bar n$,
$[\mathcal{O}_n(x=0,z_1,z_2)]$ and $[\mathcal{O}_{\bar n}(x,w_1,w_2)]$, respectively, satisfies
the following equation: 
\begin{align}\label{ONN}
\left(2(n\bar n) S_+^{(z)}-\frac12 x^2(\bar n \partial_x)\right)\langle[\mathcal{O}_n(z_1,z_2)][\mathcal{O}_{\bar n}(x,w_1,w_2)]\rangle=0\,.
\end{align}
The superscript $S_+^{(z)}$ indicates
that it is a differential operator acting on $z_1,z_2$ coordinates and we also assume  that $(x\bar n)=(x n)=0$.
The explicit expression for $S_+$  can be derived from the conformal Ward identity for the corresponding
correlator
\begin{align}\label{OOS}
\langle{\delta_+ S_R\, [\mathcal{O}^{(n)}](z)[\mathcal{O}^{(\bar n)}](x,\!w)}\rangle
=
\langle{\delta_+[\mathcal{O}^{(n)}](z)[\mathcal{O}^{(\bar n)}](x,\!w)}\rangle
\!+\!\langle{[\mathcal{O}^{(n)}](z)\,\delta_+[\mathcal{O}^{(\bar n)}](x,\!w)}\rangle,
\end{align}
bringing it to the form~(\ref{ONN}). 
Here $\delta_+$ is the transformation generated by the generator $\mathbf{K}_{\bar n}=(\bar n \mathbf{K})$ and
\begin{align}\label{KS}
\delta_+ S_R =4\epsilon\!\int\! d^dx\,
(x\bar n) (\mathcal{L}_A\!+\!\mathcal{L}_\xi\!+\!\mathcal{L}_{ghost})
+2(d\!-\!2)\bar n^\mu\!\int\! d^dx \Big(Z_c^2\,\bar c \,D_\mu c-\frac1\xi A_\mu (\partial A)\Big).
\end{align}
 Details of the calculation can be found in
Refs.~[\refcite{Braun:2013tva,Braun:2014vba}]. 
We stress that considering the correlator of two light-ray operators instead of the Green function of 
the light-ray operator with quark and antiquark fields considerably simplifies the analysis.
Indeed, the Green function is gauge-dependent and does not transform in a proper way under conformal transformations.
Another advantage is that the last term in Eq.~(\ref{KS}) which does not vanish in $d=4$ dimension and explicitly breaks
conformal symmetry of QCD Lagrangian, drops out from the correlator of gauge-invariant objects as it is reduced to
a BRST variation.

%%%%%%%%%%%%%%%%%%%%%%%%%%%%%%%%%%%%%%%%%%%%%%%%%%%%%%%%%%%%%%%%%%%%%%%%%%%%%%%%%%%%%%%%%%%%%%%%%%%%%%%%%%%%%%%%%%%%%%
\section{Two loop kernels}
%%%%%%%%%%%%%%%%%%%%%%%%%%%%%%%%%%%%%%%%%%%%%%%%%%%%%%%%%%%%%%%%%%%%%%%%%%%%%%%%%%%%%%%%%%%%%%%%%%%%%%%%%%%%%%%%%%%%%%
The two-loop kernel $h^{(2)}(\alpha,\beta)$ contains contributions of two color structures and a term proportional to the
QCD beta function,
\begin{align}\label{CACFH}
  h^{(2)}(\alpha,\beta) =8 C_F^2  h^{(2)}_1(\alpha,\beta) +4 C_F C_A h^{(2)}_2(\alpha,\beta) +4 b_0 C_F h^{(2)}_3(\alpha,\beta)\,.
\end{align}
Their noninvariant parts 
%of the kernels (\ref{CACFH}) 
can be restored from
the commutator relation~Eq.~(\ref{nest}):
\begin{align}
{}[S_+^{(0)},\mathbb{H}^{(2)}] = [\mathbb{H}^{(1)},\Delta S_+^{(1)}]\,.
\end{align}
Note that $\Delta S_+^{(1)}$ (\ref{exactS}) contains terms in $b_0$ and $C_F$.
Hence the commutator $[\Delta S_+^{(1)},\mathbb{H}^{(1)}]$ contains two color structures only, $b_0 C_F$ and $C_F^2$,
respectively.
It follows that the kernel $ h^{(2)}_2(\alpha,\beta)$~(\ref{CACFH}) satisfies the homogeneous
equation $[S_+^{(0)},\mathbb{H}_2^{(2)}] = 0 $, alias it is $SL(2)$-invariant and can be written as a function of the
conformal ratio, $h^{(2)}_2(\alpha,\beta)= h^{(2)}_2(\tau)$.

Going through the calculations one gets~\cite{Braun:2014vba}
\begin{align}
h^{(2)}_1(\alpha,\beta)&=\varphi_1(\alpha,\!\beta)\!-\!\delta_+(\tau)\Big[\phi_1(\alpha)\!+\!\phi_1(\beta)\Big]
%\!+\!\varphi_1(\alpha,\!\beta)
\!+\! \theta(\bar\tau)\left[2\Li_2(\tau)\!+\!\ln^2\bar\tau\!+\!\ln\tau\!-\!\frac{1\!+\!\bar\tau}{\tau}\ln\bar\tau\right]
\notag\\
&\quad+
\theta(-\bar\tau)\left[\ln^2(-\bar\tau/\tau)-\frac2\tau\ln(-\bar\tau/\tau)\right]
+\left[-6\zeta(3)+\frac13\pi^2+\frac{21}8\right]\delta(\alpha)\delta(\beta)\,,
\notag\\
h^{(2)}_2(\alpha,\beta)&=\frac13\left({\pi^2}-4\right)\delta_+(\tau) - 2\theta(\bar\tau)\left[\Li_2(\tau)-\Li_2(1)+\frac12\ln^2\bar\tau
-\frac1\tau\ln\bar\tau+\frac53\right]
\notag\\
&\quad -\theta(-\bar\tau)\left[\ln^2(-\bar\tau/\tau)-\frac2\tau\ln(-\bar\tau/\tau)\right]
+ \left(6\zeta(3)-\frac23\pi^2+\frac{13}6\right)\delta(\alpha)\delta(\beta)\,,
\notag\\
h^{(2)}_3(\alpha,\beta)&=-\delta_+(\tau)\left[\ln\bar\alpha+\ln\bar\beta+\frac53\right]-\theta(\bar\tau)
\left[\ln(1-\alpha-\beta)+\frac{11}3\right]+\frac{13}{12}\delta(\alpha)\delta(\beta)\,.
\label{final}
\end{align}
where $\bar\tau = 1-\tau$, and the functions $\phi_1(\alpha)$ and $\varphi_1(\alpha,\beta)$ are given by the
following expressions
\begin{align}\label{phi1}
\phi_1(\alpha)&=-\ln\bar\alpha\left[\frac32-\ln\bar\alpha
+\frac{1+\bar\alpha}{\bar\alpha}\ln\alpha\right], \qquad% \phi_3(\alpha) =\ln\bar\alpha\,,
\notag\\
\varphi_1(\alpha,\beta)&
=-\theta(1-\tau)\Big[\frac12\ln^2(1-\alpha-\beta)+\frac12\ln^2\bar\alpha+\frac12\ln^2\bar\beta-\ln\alpha\ln\bar\alpha-\ln\beta\ln\bar\beta
\notag\\
%\label{varphi1}
&\quad-\frac12\ln\alpha-\frac12\ln\beta
+\frac{\bar\alpha}\alpha\ln\bar\alpha+\frac{\bar\beta}\beta\ln\bar\beta\Big],
%+\bar\varphi_1(\tau)\,.
%\notag\\
%\varphi_3(\alpha,\beta)&=-\ln(1-\alpha-\beta)\theta(1-\tau)\,.
%%+\bar\varphi_3(\tau)\,.
\end{align}
%%%%%%%%%%%%%%%%%%%%%%%%%%%%%%%%%%%%%%%%%%%%%%%%%%%%%%%%%%%%%%%%%%%%%%%%%%%%%%%%%%%%%%%%%%%%%%%%%%%%%%%%%%%%%%%%%%%%%%
\section{Conclusion}
%%%%%%%%%%%%%%%%%%%%%%%%%%%%%%%%%%%%%%%%%%%%%%%%%%%%%%%%%%%%%%%%%%%%%%%%%%%%%%%%%%%%%%%%%%%%%%%%%%%%%%%%%%%%%%%%%%%%%%
Our result for the two-loop evolution kernels of flavor-nonsinglet operators in Eqs.~(\ref{CACFH}), (\ref{final})
is equivalent to the corresponding evolution equation  for GPDs obtained
in Ref.~[\refcite{Belitsky:1999hf}] in momentum space and has manifest $SL(2)$-symmetry properties.
This feature presents the crucial advantage of the
light-ray operator formalism which makes this technique attractive for
higher-order calculations. Exact conformal symmetry of QCD the critical point is very helpful on
intermediate steps of the calculation as it provides one with algebraic group-theory methods to calculate
the commutators of integral operators that appear in Eqs.~(\ref{nest}).
Evolution equations for GPDs can be obtained from our expressions by a Fourier
transformation which is rather straightforward, cf.~[\refcite{Braun:2009mi}].

%%%%%%%%%%%%%%%%%%%%%%%%%%%%%%%%%%%%%%%%%%%%%%%%%%%%%%%%%%%%%%%%%%%%%%%%%%%%%%%%%%%%%%%%%%%%%%%%%%%%%%%%%%%%%%%%%%%%%%

\end{document}